\documentclass[runningheads]{llncs}
\usepackage[T1]{fontenc}
\usepackage{graphicx}
\usepackage{amsmath}
\usepackage{booktabs}
\usepackage{pifont}

\usepackage{tabularx}

\usepackage{longtable}

\usepackage{xcolor}

\usepackage{subcaption}

\newcommand{\cmark}{\ding{52}}  
\newcommand{\xmark}{\ding{55}}  
\newcommand{\warn}{\ding{222}}

\usepackage{cite}       
\usepackage[colorlinks=true, linkcolor=blue, citecolor=blue, urlcolor=blue]{hyperref}  
\usepackage{orcidlink}

\usepackage{marvosym} 
\usepackage{pifont}
\begin{document}
\title{QONNECT: A QoS-Aware Orchestration System for Distributed Kubernetes Clusters}
\titlerunning{QONNECT: A QoS-Aware Orchestration System}

\author{Haci Ismail Aslan\textsuperscript{(\Letter)}\orcidlink{0000-0002-3647-5054} \and 
Syed Muhammad Mahmudul Haque\orcidlink{0009-0000-0801-5432} \and
Joel Witzke\orcidlink{0000-0002-0831-8078} \and
Odej Kao\orcidlink{0000-0001-6454-6799}
}

\authorrunning{H. I. Aslan et al.}
%
\institute{Technische Universität Berlin, Berlin, Germany\\ \email{\{aslan, joel.witzke, odej.kao\}@tu-berlin.de}, \\ \email{haque@campus.tu-berlin.de}}

\maketitle              
\begin{abstract}

Modern applications increasingly span across cloud, fog, and edge environments, demanding orchestration systems that can adapt to diverse deployment contexts while meeting Quality-of-Service (QoS) requirements. Standard Kubernetes schedulers do not account for user-defined objectives such as energy efficiency, cost optimization, and global performance, often leaving operators to make manual, cluster-by-cluster placement decisions. To address this need, we present QONNECT, a vendor-agnostic orchestration framework that enables declarative, QoS-driven application deployment across heterogeneous Kubernetes and K3s clusters. QONNECT introduces a distributed architecture composed of a central Knowledge Base, Raft-replicated Resource Lead Agents, and lightweight Resource Agents in each cluster. Through a minimal YAML-based interface, users specify high-level QoS goals, which the system translates into concrete placement and migration actions. Our implementation is evaluated on a federated testbed of up to nine cloud–fog–edge clusters using the Istio Bookinfo microservice application. The system demonstrates dynamic, policy-driven microservice placement, automated failover, QoS-compliant rescheduling, and leader re-election after node failure—all without manual intervention. By bridging the gap between declarative deployment models and operational QoS goals, QONNECT transforms the cloud–edge continuum into a unified, self-optimizing platform.

\end{abstract}
\section{Introduction}

The cloud–edge continuum has emerged as a response to the varying latency, privacy, and availability requirements of modern distributed applications \cite{2011TheNISTDefinitionofCloudComputing,2022CloudContinuumTheDefinition}. While cloud computing offers elasticity, scalability, and cost efficiency, its centralized model remains insufficient for latency-sensitive or connectivity-constrained scenarios, particularly at the mobile and edge levels \cite{2017FogComputing,2013StudyonAdvantagesandDisadvantagesofCloudComputingTheAdvantagesofTelemetryApplicationsintheCloud}. Fog computing mitigates these limitations by introducing intermediate layers that bring compute and storage closer to data sources \cite{2017FogComputing,2018FogComputingConceptualModel,2018AResearchPerspectiveonFogComputing}, but managing large-scale, heterogeneous fog infrastructures poses challenges in energy efficiency, security, and fault tolerance \cite{2017FogComputingIssuesChallengesAndFutureDirections,2018AResearchPerspectiveonFogComputing,2016AnOverviewOfFogComputingAndItsSecurityIssues}. Edge computing extends this trend further, enabling near-device processing to support real-time workloads, yet it introduces its own constraints in terms of capacity, reliability, and exposure to distributed attack surfaces \cite{2020AProposedComputationWhichBenefitsFromTheCooperationOfDewEdgeAndCloudComputations,2024DynamicTrustSecurityApproachForEdgeComputingBasedMobileIoTDevicesUsingArtificialIntelligence,2022SecurityProvisionsInSmartEdgeComputingDevicesUsingBlockchainAndMachineLearning}.

\begin{figure}
  \centering
  \includegraphics[width=0.7\linewidth]{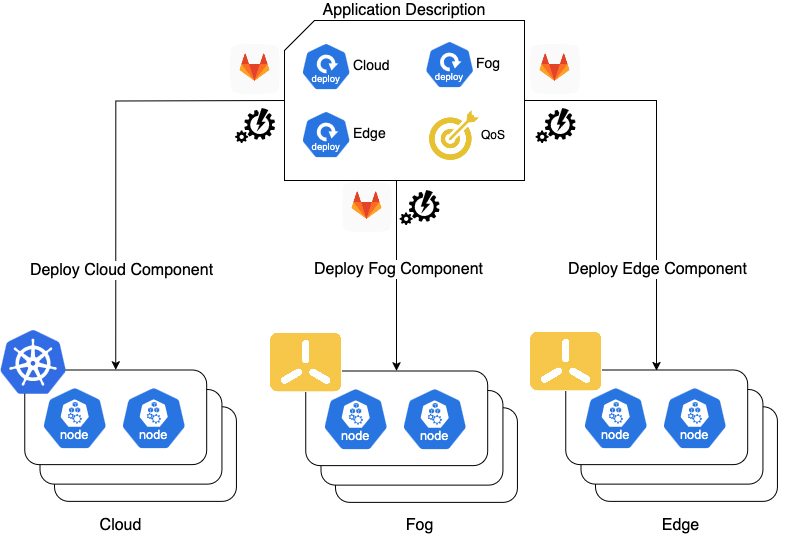}
  \caption{Target approach.}
  \label{fig:target-deploy}
\end{figure}

Across the cloud–fog–edge continuum, no single model is universally optimal. Instead, hybrid approaches that coordinate centralized and decentralized resources offer better flexibility and performance \cite{hybrid}. Cloud-native orchestration tools like Kubernetes (K8s) have enabled scalable, infrastructure-agnostic application management in cloud environments \cite{BorgOmegaK8s}. Lightweight variants such as k3s and KubeEdge extend orchestration to fog and edge settings \cite{CloudNativeWorkloadOrchestrationEdgeReviewFutureDirections}. However, existing tools fall short in managing distributed, multi-layered applications with diverse QoS constraints \cite{2021MiCADOEdge,QoSAwareOrchestrationCloudEdgeContinuumApp}.

A robust cloud-to-edge orchestrator must meet seven key criteria \cite{2021MiCADOEdge}: (1) unified deployment across layers, (2) technology agnosticism, (3) multi-cloud capability, (4) dynamic runtime optimization, (5) extensible policy definitions, (6) support for both containers and virtual machines (VMs), and (7) production-grade robustness and security. Kubernetes partially satisfies these through autoscaling, extensibility, and VM integration, but lacks native support for QoS-aware scheduling and cross-cluster placement optimization.

While declarative tools like Helm and Kustomize simplify multi-service deployment, placing services based on latency, cost, or energy remains largely manual. The default Kubernetes scheduler is unaware of such QoS goals, requiring administrators to hardcode node placement and manually update CI/CD pipelines during initial deployment or disaster recovery \cite{2022Nautilus}.

To overcome these limitations, we propose QONNECT, an orchestration framework that automates the deployment of interconnected microservices across cloud, fog, and edge environments. Our solution aims to fulfill the key orchestrator requirements while minimizing manual intervention, guided by initiatives such as Swarmchestrate \cite{swarm,swarm-v2} and informed by real-world constraints. The envisioned approach is illustrated in Figure~\ref{fig:target-deploy}. We summarize our main contributions as follows:

\begin{itemize}
  \item \textbf{Vendor‑agnostic architecture.}  
        A clean separation of global state (Knowledge Base), distributed decision-making (Raft‑replicated RLAs), and local actuation (RAs) keeps the framework independent of cloud‑provider APIs.

  \item \textbf{QoS‑aware scheduler.}  
        A weighted Borda count translates declarative vectors for energy, cost, and performance into concrete placements, requiring only a minimal YAML addition to existing manifests; the ranking component can be swapped for alternative algorithms.
        
  \item \textbf{Fault‑tolerant control plane.}  
        RLAs form a Raft quorum, automatically elect new leaders, and cooperate with RAs for transparent workload migration.
\end{itemize}

\section{Background}

\subsection{Kubernetes and Lightweight Distributions}

Kubernetes is the de facto standard for orchestrating containerized applications at scale. It decouples application deployment from infrastructure management using a declarative model driven by a central API server \cite{Poulton_2021}. A typical cluster consists of a control plane (API server, scheduler, etcd, controller managers) and worker nodes that run containerized workloads grouped into pods. Deployments and ReplicaSets enable replication and lifecycle management. Kubernetes offers autoscaling, service discovery, self-healing, and declarative management of networking, storage, configuration, and secrets—supporting stateless, dynamic applications. To extend these features to resource-constrained environments, lightweight distributions like k3s bundle components and replace etcd with SQLite. Designed for edge and fog deployments, k3s supports Helm, CRDs, and runtimes like containerd, making it ideal for low-power or remote nodes.



\subsection{Consensus and Orchestration}

\begin{figure}[h]
\vspace{-5mm}
  \centering
  \begin{subfigure}[b]{0.5\linewidth}
    \centering
    \includegraphics[width=\linewidth]{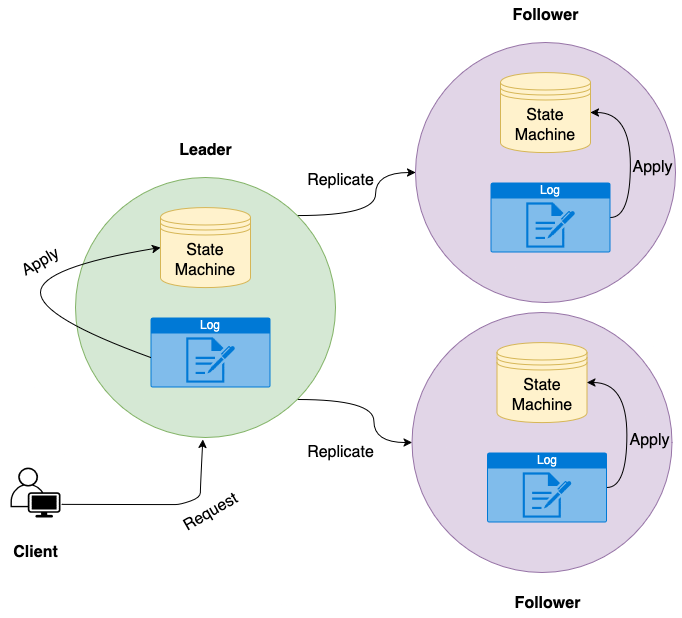}
    \caption{Replicated State Machine (RSM) model.}
    \label{fig:rsm}
  \end{subfigure}
  \hfill
  \begin{subfigure}[b]{0.7\linewidth}
    \centering
    \includegraphics[width=\linewidth]{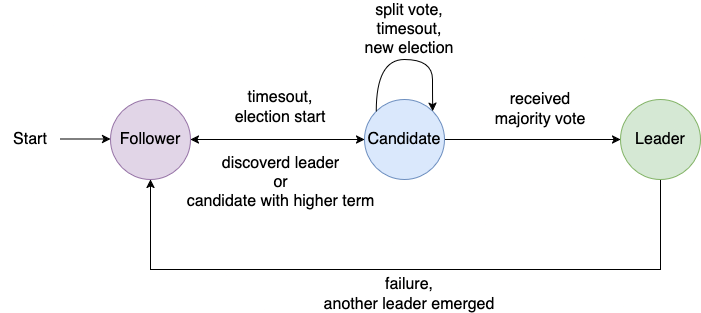}
    \caption{Raft node states and transitions.}
    \label{fig:raft-node-states}
  \end{subfigure}
  \caption{Overview of Raft-based architecture components.}
  \label{fig:raft-overview}
  \vspace{-5mm}
\end{figure}

Reliable orchestration across distributed clusters requires strong consistency guarantees for a shared state. One commonly adopted \cite{RSM_use2,book_RSM} model is the \textit{Replicated State Machine (RSM)}, where commands from clients are executed in the same order on all nodes. As shown in Figure~\ref{fig:rsm}, this requires consistent log replication and fault tolerance to maintain synchronization across nodes.


\textit{Raft}~\cite{ongaro2014search} is a consensus algorithm designed to implement RSMs with both understandability and practical deployment in mind. It simplifies consensus into three distinct components: \textit{leader election}, \textit{log replication}, and \textit{safety}. A Raft cluster consists of an odd number of nodes (e.g., 3 or 5), allowing the system to tolerate up to $\lfloor n/2 \rfloor$ node failures while maintaining quorum. Figure~\ref{fig:raft-node-states} illustrates the finite-state machine that governs each node in Raft. At any time, a node is in one of three states: \textit{leader}, \textit{follower}, or \textit{candidate}. The leader handles client requests, appends them to its log, and replicates them to followers. Followers remain passive, responding only to messages from the leader or candidates.


When a leader fails or becomes unreachable, followers won't receive its heartbeat and will consequently transition to candidates, starting a new election. A majority vote elects a new leader, who then resumes log replication. Log consistency is maintained by matching log indices and terms, ensuring that committed entries are preserved across leadership changes.

Raft's simplicity and robustness make it an ideal choice for managing distributed configuration states in orchestrators, particularly for multicluster coordination, leader tracking, or implementing consistent control logic across cloud\-fog-edge environments.

\subsection{QoS-Aware Scheduling and Borda Count}

Traditional Kubernetes schedulers assign pods based on resource availability and node affinity, but lack native support for QoS-based policies (e.g., latency, cost, energy). To introduce such logic, external schedulers or controllers can rank candidate nodes using preference models. 

One such model is the \textit{Borda Count}, a voting method where each alternative (e.g., node) is assigned a score based on ranked preferences from multiple criteria. This allows fine-grained, weighted decisions for scheduling microservices with complex placement needs.

\section{Architecture}

OONNECT is a distributed orchestration framework designed to enable Quality-of-Service (QoS)-aware application deployment across heterogeneous cloud, fog, and edge environments. It abstracts away the complexity of coordinating multiple Kubernetes and K3s clusters, allowing operators and developers to specify high-level QoS goals through declarative policies. The system autonomously handles microservice placement, migration, and failover, adapting to dynamic infrastructure conditions while maintaining compliance with QoS constraints.

\subsection{Requirements}

The primary objective is to facilitate the cross-domain deployment and management of application components based on the user's QoS goals. This includes energy efficiency, cost and performance considerations (e.g., CPU, memory, storage, and bandwidth). Furthermore, the system must be capable of redeploying components upon failures or disruptions, ensuring that user applications continue to run uninterrupted. In summary, the framework must provide functional (FR) and non-functional (NFR) requirements such as:

\begin{itemize}
    \item Flexible, user-directed domain placement (e.g., cloud, fog, or edge) without requiring manual setup.

    \item Adherence to user-defined goals for energy efficiency, cost, and application performance.

    \item Redeployment mechanisms when disruptions occur, maintaining the same QoS standards.

    \item Overall fault tolerance to handle unexpected events or resource contention.
\end{itemize}

By satisfying these functional and non-functional requirements, the proposed system will ensure reliable, high-performance, cost-effective, and energy-efficient deployments across cloud, fog, and edge domains .

\subsubsection{User-Guided Domain Deployment (FR1)} \label{fr1} The system shall deploy each application component to the domain cluster (cloud, fog, or edge) as specified by the user based on operational and QoS requirements. The actual scheduling and instantiation of components in the user-selected domain shall be handled swiftly and transparently.

\subsubsection{QoS-Driven Deployment (FR2)}\label{fr2} The system shall deploy application components based on user-defined QoS goals, ensuring optimal alignment with specified requirements such as energy consumption, operational cost, and performance. This includes:

\vspace{-2.5mm}
\begin{enumerate}
    \item QoS-Aware Scheduling: Selecting deployment locations by evaluating available resources against user QoS targets (e.g., cost, energy, and performance).
    
    \item Dynamic Adaptation: Adjusting deployment decisions dynamically if the QoS parameters or resource availability change. 
\end{enumerate}

\vspace{-5.0mm}
\subsubsection{Disruption Recovery and Redeployment (FR3)}\label{fr3} The system shall detect any disruptions (e.g., node failures or resource overloads) affecting running application components and automatically redeploy them to another cluster within the same domain if needed. This requirement involves:

\vspace{-2.5mm}
\begin{enumerate}
    \item Continuous Monitoring: Tracking cluster health, resource usage, and application status to identify problems promptly.

    \item Automatic Migration: Re-deploying affected components in alternative clusters according to the user's QoS configurations and preferences.
\end{enumerate}

\vspace{-5.0mm}
\subsubsection{Fault Tolerance (NFR1)} \label{nfr1}

The framework shall exhibit fault-tolerant behavior, continuing to operate despite failures in hardware, software, or network resources. This involves employing backup instances to mitigate single points of failure.

\subsection{Concept}

\begin{figure}[h]
  \centering
  \includegraphics[width=\linewidth]{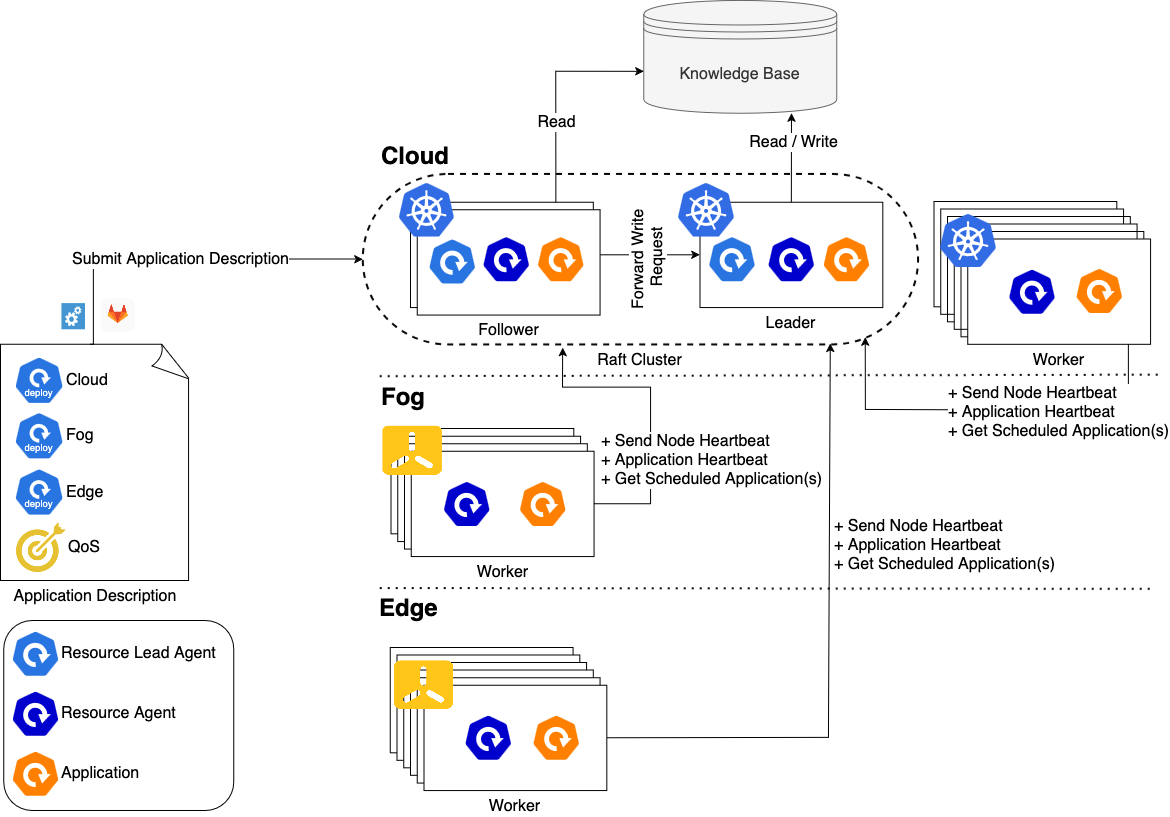}
  \caption{Architecture overview: component interaction across cloud, fog, and edge domains.}
  \label{fig:arch}
\end{figure}

The QONNECT architecture is influenced by the principles of \textit{Clean Architecture}\footnote{\url{https://blog.cleancoder.com/uncle-bob/2012/08/13/the-clean-architecture.html}; last accessed: \today}, which promote a clear separation of concerns across domain logic, use cases, and interface boundaries. This design choice enhances modularity, testability, and portability across diverse deployment environments. As illustrated in Figure~\ref{fig:arch}, QONNECT consists of three core components: the \textit{Knowledge Base (KB)}, the \textit{Resource Lead Agent (RLA)}, and the \textit{Resource Agent (RA)}—each fulfilling distinct roles in orchestrating services across distributed domains.

Each domain manages its own cluster(s). Cloud clusters typically run full Kubernetes; edge clusters use lightweight k3s; and fog clusters may use either, depending on available resources. Clusters consist of a control-plane and worker nodes. Worker nodes host application components and are monitored by a cluster-local RA, while at least one cluster on cloud includes an RLA to enable leadership and coordination.

The \textit{Knowledge Base (KB)} acts as a global metadata repository for all clusters, nodes, and application deployments. It holds information such as available resources, node health, workload status, and user QoS requirements. While logically centralized, the KB can be backed by any reliable distributed storage (e.g., replicated SQL/NoSQL databases or blockchain-based solutions), ensuring resilience and fault tolerance.

The \textit{Resource Lead Agent (RLA)} is responsible for cluster coordination, system-wide scheduling, and consensus management. RLAs run on selected nodes and collectively form a Raft quorum for leader election and consistent decision-making. The elected RLA leader becomes the system's API entry point, serving orchestration requests and issuing deployment decisions based on user goals (e.g., minimizing cost, reducing latency, or conserving energy). The RLA evaluates all registered clusters against user-defined QoS constraints and dynamically selects the most suitable ones.

The \textit{Resource Agent (RA)} is deployed in every application-hosting cluster. It interacts with the local Kubernetes or k3s API to receive and apply deployment instructions issued by the RLA. The RA also collects real-time telemetry from the cluster—covering node status, application health, and resource utilization—and forwards this data to the RLA. This ensures the orchestration loop remains aware of environmental changes and can trigger redeployments when necessary.

Each RA is stateless and reactive, ensuring fault-tolerant behavior with minimal footprint. RAs also act as execution proxies, translating higher-level scheduling decisions into native Kubernetes manifests and ensuring that inter-service communication and dependency configurations are honored locally. The entire system is designed to support QoS-aware deployments while ensuring resilience and minimal manual intervention. If a cluster or node becomes unavailable, the RLA detects the failure and initiates a migration of the affected application to another suitable cluster within the same or a different domain, maintaining compliance with user goals.


By combining centralized decision-making (via the KB and RLA leader) with decentralized enforcement (via RAs), the architecture achieves scalable, cross-domain orchestration suited to dynamic, resource-constrained environments across the cloud–fog–edge continuum.

\subsection{Implementation}

This section details our system's implementation, justifying the claimed achievements. Our implementation\footnote{\url{https://github.com/dos-group/QONNECT}; last-accessed: \today} is open-sourced anonymously. 

\paragraph{Raft} Instead of a custom Raft implementation, we forked \lq etcd/contrib/raftexample\rq , retaining its bare-bones consensus layer (state machine, Write Ahead Log (WAL), snapshot) and replacing its key-value store/transport with a minimalist HTTP-based REST transport.

\paragraph{Scheduler} The leader RLA in a Raft cluster serves the scheduler and the scheduler periodically:
\begin{enumerate}
    \item Places pending services by assigning clusters/nodes using the Borda scorer, persisting decisions in the KB.
    \item Re-queues stalled services; workloads with heartbeats older than a grace period are reset.
\end{enumerate}

The Borda scorer's `ScoreAndFilterNodes' method receives a node snapshot and QoS, proceeding via these steps:
\begin{enumerate}
    \item Eligibility filter: Removes unready, unschedulable, or resource-pressured nodes; an empty result stops processing.
    \item Per-attribute Borda ranking: Sorts slices and assigns Borda scores for energy, pricing, and capacity (CPU, memory, bandwidth, storage); lower values win for energy/pricing, higher for performance.
    \item Capacity factor: Sums the four capacity Bordas into a single \lq CapacityBorda\rq ~representing raw node performance.
    \item QoS-weighted node score: Calculated as
        \[
          w =  Energy\_{\text{Borda}}\cdot q_E +
               Pricing\_{\text{Borda}}\cdot q_P +
               Capacity\_{\text{Borda}}\cdot q_C ,
        \]
        where \((q_E,q_P,q_C)\) are user weights (energy, pricing, performance), defaulting to \((1,1,1)\) if all are \(0.0\).
    \item Threshold filter: Keeps nodes with score $w$ greater than or equal to the arithmetic mean of all $w$ values.
    \item Cluster aggregation: Sums node scores per cluster, ordering clusters by retained score (total score as tie-breaker).
    \item Return value: ID of the highest-ranked cluster and retained node names within it.
\end{enumerate}

\paragraph{REST API} Manifest portability across domains is enabled by two conventions: (1) Each domain manifest requires an Ingress with the application name as its first path segment, and (2) cross-domain placeholders (IP addresses for cloud-fog-edge clusters) are used instead of hard-coded addresses, which are replaced by the RA with concrete cluster IPs before manifest application. The `application' field is a thin CRD-style YAML for scheduler-required metadata (name, labels, QoS).

\subsubsection{Resource Agent}
The RA, which also follows the Clean Architecture principles, is responsible for the orchestration steps described in the following subsections.

\paragraph{Registration} On startup, the RA checks for prior registration via the `self' ConfigMap in the namespace. If present, cluster ID and role are loaded; otherwise, the cluster joins for the first time by determining its external IP, submitting it to the bootstrap RLA, and caching the returned UUID in a new ConfigMap. During bootstrap, auxiliary ConfigMaps for cluster IP mapping and running applications are also verified for subsequent operations.

\paragraph{Send node snapshot} The RA uses the Kubernetes API to enumerate, filter control-plane nodes, and send the node snapshot to the RLA.

\paragraph{Polls cluster config} The RA periodically fetches cluster ID $\rightarrow$ ingress IP mapping from the RLA via REST, updating the local ConfigMap for application deployment.

\paragraph{Polls applications} The RA retrieves recently scheduled workloads from the RLA's dedicated REST endpoint. For each, it: (1) ensures/creates the target namespace, (2) applies Kubernetes objects, merging labels, and replacing cross-domain placeholders with concrete IPs from the cluster config, and (3) restricts deployments to scheduler-chosen worker nodes. On successful deployment, the RA adds/updates the applications ConfigMap with application ID, version, applied objects, and timestamp, serving as the single source of truth for health checks and documenting assigned workloads.

\paragraph{Send application heartbeat} For each application in the applications ConfigMap, the RA reports status to the RLA: (1) healthy (all Deployments reached desired replica count), (2) progressing (at least one Deployment rolling out, within timeout), or (3) failed (e.g., crash-loop, missing objects, timeout, requiring manual intervention).

\paragraph{Cleanup application} If the RLA responds to an application heartbeat with `404 Not Found', the RA concludes workload withdrawal, immediately deleting the Kubernetes namespace (removing deployed objects) and its ConfigMap entry, preventing future heartbeat cycles.

\section{Evaluation}

\subsection{Setup}
For evaluation, we established a testbed based on Kind\footnote{\url{https://kind.sigs.k8s.io}; last-accessed: \today} comprising nine clusters: three each for cloud, fog, and edge. Within each domain, clusters were assigned one of three profiles: energy-efficient, cost-efficient, or performance-efficient. Each cluster consists of one control-plane and two worker nodes. As clusters run in Docker on a single machine, real telemetry is unavailable; thus, representative energy, cost, and bandwidth values were manually assigned based on public data.

\subsubsection{Energy Consumption} Energy consumption was calculated using coefficients from Cloud Carbon Footprint's Appendix I\footnote{\url{https://www.cloudcarbonfootprint.org/docs/methodology/\#appendix-i-energy-coefficients}} and the formula: $E = ((W_{\mathrm{idle}} + W_{\mathrm{max}})/2) \times \mathrm{PUE} \times 1h / 1000$, where $\mathrm{PUE}$ is power usage effectiveness and unitless. This yielded $E_{\mathrm{AWS}} \approx 0.0024042\ \mathrm{kWh}$  and $E_{\mathrm{GCP}} \approx 0.0027335\ \mathrm{kWh}$. We assigned the lowest, midpoint ($E_{\mathrm{mid}} \approx 0.0025689\ \mathrm{kWh}$), and highest values to energy-efficient, cost-efficient, and performance-efficient profiles, respectively.

\subsubsection{Cost (EUR/h)} Cost values, based on AWS On-Demand pricing\footnote{\url{https://aws.amazon.com/de/ec2/pricing/on-demand/} (Location Type: AWS Region, Region: US East (Ohio), OS: Linux)} (assuming 1 EUR = 1 USD), were set as: $C_{\min} = 0.0042\ \mathrm{EUR/h}$ (\texttt{t4g.nano}) and $C_{\max} = 32.7726\ \mathrm{EUR/h}$ (\texttt{p4d.24xlarge}). These were assigned to cost-efficient, energy-efficient ($C_{\mathrm{mid}} \approx 16.3884\ \mathrm{EUR/h}$), and performance-efficient profiles, respectively.

\subsubsection{Network Bandwidth} Network bandwidth values were derived from AWS specs (5 Gbps for \texttt{t4g.nano}, capped at 100 Gbps for \texttt{p4d.24xlarge} from 400 Gbps). We assigned 5 Gbps to cost-efficient, 52.5 Gbps ($B_{\mathrm{mid}}$) to energy-efficient, and 100 Gbps to performance-efficient profiles.

The assigned parameters for each cluster profile are summarized in Table~\ref{tab:setup-params}. We applied these same values to fog and edge clusters for consistency, acknowledging that real-world values would differ but maintaining clear distinctions across profiles. To simulate energy- and cost-efficient characteristics, we manually reduced reserved system resources on their worker nodes (5 GiB memory, 4 CPU cores, 2 GiB ephemeral storage) via \texttt{kubeadmConfigPatches} during setup.

\begin{table}[ht]
  \centering
  \caption{Assigned energy, cost, and bandwidth parameters for each cluster profile.}
  \label{tab:setup-params}
  \begin{tabular}{lrrr}
    \toprule
    \textbf{Cluster Profile}      & \textbf{Energy (kWh)} & \textbf{Cost (EUR/h)} & \textbf{Bandwidth (Gbps)} \\
    \midrule
    Energy-efficient             & 0.0024042             & 16.3884               & 52.5                      \\
    Cost-efficient               & 0.0025689             & 0.0042             & 5                         \\
    Performance-efficient        & 0.0027335             & 32.7726               & 100                       \\
    \bottomrule
  \end{tabular}
  \vspace{-5.0mm}
\end{table}

\subsubsection{Test application}

For system evaluation, we adapted the Istio Bookinfo microservices application\footnote{\url{https://istio.io/latest/docs/examples/bookinfo/noistio.svg}; last-accessed: \today}, a multi-service e-commerce system, with minimal configuration modifications and added a Kubernetes Ingress. The Bookinfo application comprises \texttt{productpage}, \texttt{details}, \texttt{reviews}, and \texttt{ratings} services. To simulate a cloud–fog–edge architecture, we deployed \texttt{productpage} on the cloud cluster, \texttt{details} and \texttt{reviews} on the fog cluster, and \texttt{ratings} on the edge cluster. Our deployment retained the overall structure but redistributed services to align with our experimental setup.

\subsection{Results}

To test the systems requirements we conducted a series of tests using the developed QONNECT prototype:
\begin{enumerate}
  \item Deploy the test application prioritizing performance. Here QONNECT should schedule the application's components to the `performance' clusters in each domain.
  \item After initial deployment, we change the applications QoS requirement to prioritize energy efficiency instead. The components of the application should be redeployed to the corresponding `energy' clusters.
  \item After initial deployment, we simulate a cluster failing by deleting the RA within the `performance' cluster of the edge domain. QONNECT should react accordingly be redeploying the ratings component to a different cluster within the edge domain.
  \item Test the resilience against lead agent failure by deleting the RLA currently acting as raft leader. In our case that was the RLA of the `energy' cluster.
\end{enumerate}

\begin{figure}
  \centering
  \includegraphics[width=0.7\linewidth]{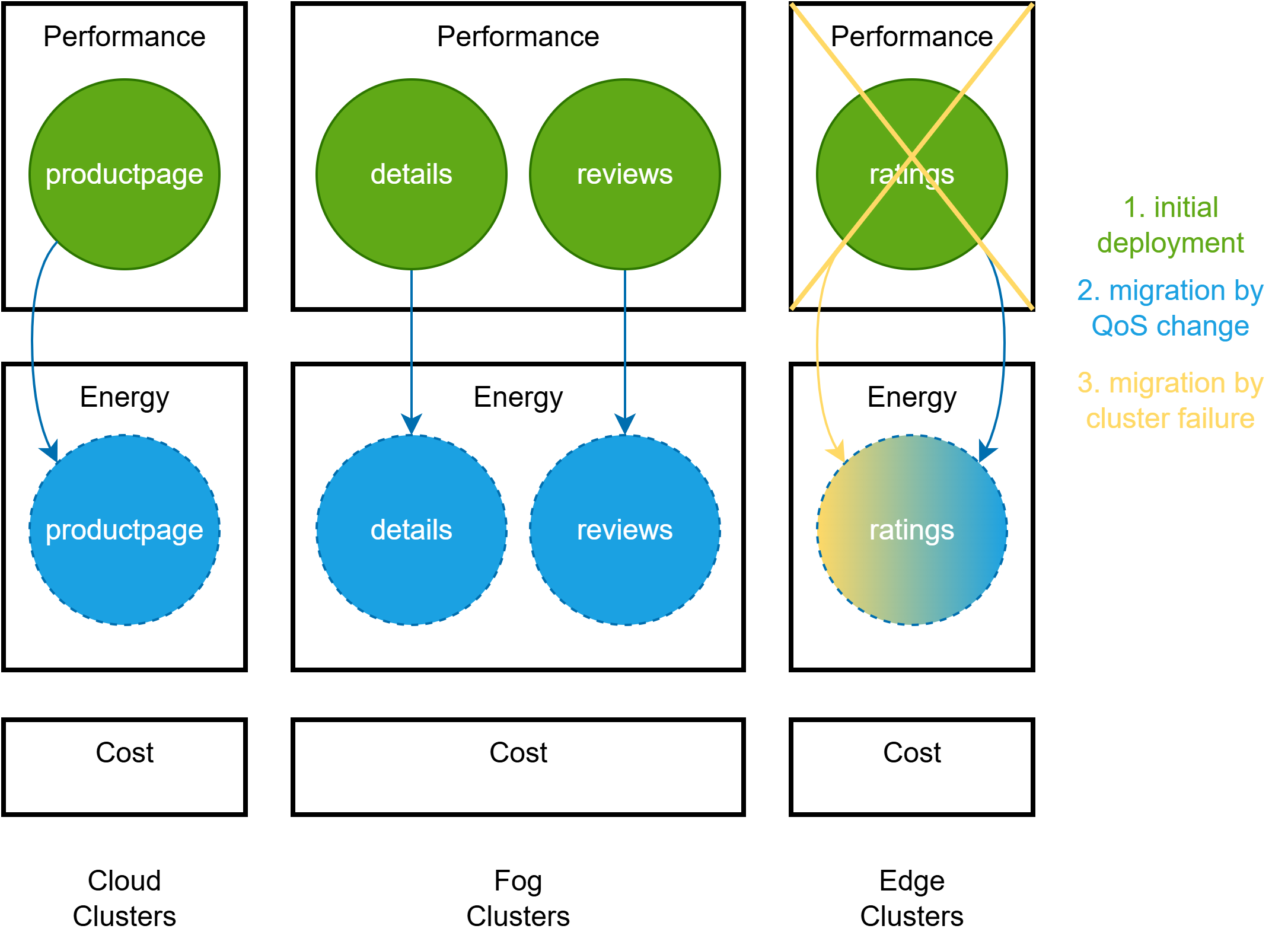}
  \caption{QONNECT prototype behavior in tests 1 to 3. The color specifies the corresponding test case.}
  \label{fig:evaluations}
\end{figure}

Figure~\ref{fig:evaluations} shows the anticipated---and achieved---outcomes from executing the tests 1, 2 and 3. Test 4 also finished successfully, electing the RLA in the `performance' cluster as new leader.
Scripts conducting these tests can be found in the same repository together with our QONNECT prototype\footnote{\url{https://github.com/dos-group/QONNECT/blob/main/resource-lead-agent/bruno-collection/swarmchestrate.json}; last-accessed: \today}.

In addition to experimental validation, we provide a comparative analysis of QONNECT against existing orchestration frameworks in terms of architectural and functional capabilities. As summarized in Table~\ref{tab:comparison}, QONNECT combines QoS-aware scheduling, cloud–edge continuum support, distributed scheduling, and fault tolerance within a unified framework. This highlights its potential to address limitations observed in prior systems and demonstrates its practical viability in diverse deployment environments.


\begin{table*}[ht!] 
  \centering
  \caption{Comparison of orchestration systems.}
  \label{tab:comparison} 
  \small 
  \begin{tabularx}{\textwidth}{@{} >{\raggedright\arraybackslash}X *{5}{>{\centering\arraybackslash}X} @{}}
    \toprule
    \textbf{Feature} &
    \textbf{QONNECT} &
    \textbf{MiCADO} &
    \textbf{ACOA} &
    \textbf{Nautilus} &
    \textbf{K8s (Vanilla)} \\
    \midrule
    \textbf{Cloud–Edge} &
    \cmark~Full support &
    \xmark~Cloud-focused; limited edge &
    \cmark~Full support &
    \warn~HPC-oriented &
    \warn~Not continuum-aware \\
    \midrule
    \textbf{QoS-aware scheduling} &
    \cmark~Supports with self-healing &
    \xmark~No QoS placement &
    \cmark~QoS-based placement &
    \warn~Resource-focused; lacks QoS logic &
    \warn~No built-in constraint \\
    \midrule
    \textbf{Distributed scheduling} &
    \cmark~Raft-based with RAs &
    \xmark~Terraform \& policy keeper &
    \cmark~Multi-scheduler model &
    \warn~K8s default control &
    \warn~Single control plane logic \\
    \midrule
    \textbf{VM \& container autoscaling} &
    \cmark~VM and container aware &
    \cmark~Supports via Terraform \& K8s &
    \warn~Focus on placement, not scaling &
    \warn~Scaling by base scheduler &
    \cmark~HPA and Cluster Autoscaler \\
    \midrule
    \textbf{Self-healing and failover} &
    \cmark~Leader election, redeployment &
    \warn~Container restarts, no migration &
    \warn~Failover not emphasized &
    \cmark~Pod recovery, monitoring &
    \cmark~Supports via add-ons \\
    \midrule
    \textbf{User-level QoS specification} &
    \cmark~Declarative YAML with QoS goals &
    \warn~Metric triggers in TOSCA &
    \cmark~Supports developer-level logic &
    \warn~No user QoS interface &
    \warn~Only resource requests/limits \\
    \midrule
    \textbf{Prototype / Testbed} &
    \cmark~K8s \& K3s clusters with Istio app &
    \cmark~Validated across cloud platforms &
    \cmark~Used in railway use-case &
    \cmark~Deployed on GPU clusters &
    \cmark~Widespread \\
    \bottomrule
  \end{tabularx}
\end{table*}

\section{Related Work}

Several orchestration frameworks have emerged to address deployment and QoS challenges across the cloud–edge continuum. Taherizadeh et al. \cite{2018Capillary} propose a capillary computing model that migrates Docker-based microservices across Edge, Fog, and Cloud tiers to maintain proximity to mobile clients and balance load. Their vehicle-mounted prototype demonstrates significant improvements in response time and latency variation. MiCADO-Edge \cite{2021MiCADOEdge,2019MiCADO} extends MiCADO with KubeEdge to orchestrate containers and VMs across heterogeneous infrastructures using TOSCA descriptors, supporting dynamic policy enforcement and multi-platform deployments in domains such as healthcare and video analytics. ACOA \cite{QoSAwareOrchestrationCloudEdgeContinuumApp} introduces a distributed, per-application multi-scheduler architecture built on Kubernetes, enabling QoS-aware placement through customizable scheduling algorithms and consideration of latency, reliability, and resource availability. Nautilus \cite{2022Nautilus} addresses dynamic runtime conditions through RL-based resource management, microservice migration, and latency-aware mapping of service graphs, achieving both QoS guarantees and resource efficiency. Complementing these system-specific efforts, Vaño et al. \cite{CloudNativeWorkloadOrchestrationEdgeReviewFutureDirections} survey the evolution of cloud-native orchestration at the edge, identifying trends in lightweight Kubernetes distributions, container runtimes, and emerging technologies such as WebAssembly-based virtualization.


\section{Conclusion and Future Work}

This work introduced QONNECT, a QoS-driven orchestrator for heterogeneous Kubernetes and K3s clusters spanning cloud to edge environments. The system enables intent-based deployment by translating high-level QoS vectors into concrete placement and migration decisions, using a modular scheduler architecture and a fault-tolerant Raft-based control plane. Through a federated nine-cluster testbed, QONNECT demonstrated its ability to maintain service-level objectives across dynamic conditions, automatically recovering from node and agent failures while preserving quorum and consistency. By demonstrating reliable, policy-driven orchestration under dynamic conditions, this work takes a step toward invisible infrastructure—where developers express what they need, and the system decides where and how to run it.

While promising, the prototype was evaluated under synthetic conditions. Future work will extend to large-scale, real-world deployments with richer QoS dimensions—especially latency—and integrate telemetry-driven energy models and predictive migration strategies.

\begin{credits}
\subsubsection{\ackname} This paper was partially supported by the Swarmchestrate project of the European Union’s Horizon 2023 Research and Innovation programme under grant agreement no. 101135012.

\subsubsection{\discintname}
The authors have no competing interests to declare that are relevant to the content of this article.
\end{credits}
%
%
%
\bibliographystyle{splncs04}
\bibliography{references.bib}

\end{document}